\documentclass[12pt]{iopart}
\usepackage{tensor}
\usepackage{iopams}
\usepackage{hyperref}

\begin{document}
\title{Conformal fluctuations do not establish a minimum length}
\author{C Cunliff}
\address{Department of Physics,
University of California-Davis, Davis, CA 95616, USA}
\ead{cunliff@physics.ucdavis.edu}

\begin{abstract}
This paper corrects an earlier work suggesting that the quantum expectation value of the proper length is bounded from below by the Planck length.  The original calculation examined fluctuations of the conformal factor of Einstein-Hilbert gravity.  However, in Einstein-Hilbert gravity, the conformal factor is not a dynamical field subject to fluctuations.  This paper performs the same calculation using the trace anomaly-induced effective action for the conformal factor and finds that, while conformal fluctuations modify the short-distance behavior of the interval, the interval still approaches zero in the coincidence limit.
\end{abstract}
\maketitle

\section{Introduction}
Nearly 30 years ago, Padamanabhan \cite{Padmanabhan1984Pl,Padmanabhan1985Ps} performed a simple calculation suggesting that quantum gravitational fluctuations place a lower bound on distance measurements.  He considered fluctuations of the conformal factor $\phi(x)$ in metrics of the form
\begin{equation}
  g\indices{_{\mu \nu}}(x) = \left(1 + \phi(x)\right)\indices{^2} \bar{g}\indices{_{\mu \nu}}(x), \label{FullMetric}
\end{equation}
while keeping the background metric $\bar{g}$ classical.  Crudely speaking, Padmanabhan argued that the conformal factor $\phi(x)$ has a Green's function that diverges as $\frac{1}{(x-x^\prime)^2}$, in such a way that $g_{\mu \nu}dx^\mu dx^\nu$ remains finite in the coincidence limit.

This calculation was part of a larger approach to quantum gravity
and quantum cosmology in which the conformal factor was treated as a
dynamical field to be quantized, while the rest of the metric was
treated as a classical field as in standard QFT.  This approach
sidesteps some of the thornier conceptual problems associated with
quantizing the metric, since conformal fluctuations preserve the
causal structure of spacetime.  However, this calculation is almost
certainly wrong. In pure Einstein-Hilbert gravity, the conformal
factor is not a dynamical
degree of freedom \cite{Kuchar1970}.  This is most clearly
seen using the York decomposition of symmetric tensors
\cite{York1973}, in which the conformal factor is determined by a
constraint equation similar to the Gauss law constraint in
electrodynamics.

To see where the argument went wrong, we must examine the path-integral
approach taken by Padmanabhan and Narlikar
\cite{Padmanabhan1983Aa,Narlikar1983Qc}.  The classical action and
path integral are
\begin{eqnarray}
  S[g] = \frac{1}{16 \pi G} \int \rmd^4 x \sqrt{-g}(R-2\Lambda) \label{EHAction}\\
  Z = \int [\mathcal{D} g] \textrm{exp}\left\{\rmi S[g]\right\}
  \label{PathIntegral}
\end{eqnarray}
In terms of the conformal factor and background metric, the action
becomes
\begin{equation}
  S[\bar{g},\phi] = \frac{1}{16\pi G}\int \rmd^4 x \sqrt{-\bar{g}}
  \left[\bar{R}(1+\phi(x))^2 - 2\Lambda (1+\phi(x))^4 - 6\phi^i\phi_i\right]
  \label{EHConformalAction}
\end{equation}
From here, the calculation proceeds in a straightforward manner.
Consider the expectation value of the interval in a (Minkowski)
vacuum state $\bar{g}_{\mu \nu} = \eta_{\mu \nu}$:
\begin{equation}
  \langle 0| \rmd s^2|0 \rangle = \langle 0|g_{\mu \nu}|0 \rangle \rmd x^\mu \rmd x^\nu = (1+\langle \phi^2(x) \rangle)\eta_{\mu
  \nu}\rmd x^\mu \rmd x^\nu.
\end{equation}
However, $\langle \phi^2 \rangle$ evaluated at a single event
diverges. Using covariant point-splitting, we instead evaluate the interval between \emph{two
events} $x^\mu$ and $y^\mu \equiv x^\mu + \rmd x^\mu$, in the limit that
$x^\mu \rightarrow y^\mu$.  With the notation $\bar{l^2} =
\eta_{\mu \nu}\rmd x^\mu \rmd x^\nu$, we examine
\begin{equation}
  \lim_{x\rightarrow y} \langle \rmd s^2 \rangle \equiv \lim_{x\rightarrow y} (1 +
  \langle\phi(x)\phi(y)\rangle)\eta_{\mu \nu}\rmd x^\mu \rmd x^\mu = \lim_{x\rightarrow y}
  (1+ \langle \phi(x)\phi(y) \rangle)\bar{l^2} \label{Interval}
\end{equation}
With $\bar{g}_{\mu \nu}=\eta_{\mu \nu}$, the action
\eref{EHConformalAction} is just the action for a massless scalar
field, albeit with a negative sign\footnote{Obtaining the clean result
\eref{EHPlanckLength} requires a nonstandard definition of the Planck
length, $L_p^2 = \frac{4\pi}{3}\frac{G}{\hbar c^3}$.}, $S[\phi]=-\frac{1}{2 L_p^2}\int
\phi^i \phi_i \rmd^4x$. The Green's
function is
\begin{equation}
  \langle \phi(x)\phi(y) \rangle = \frac{L_p^2}{4\pi^2} \cdot \frac{1}{(x-y)^2}
\end{equation}
and so the interval becomes
\begin{eqnarray}
  \lim_{x\rightarrow y} (1 + \langle \phi(x)\phi(y) \rangle)\bar{l^2}
  &=&
  \lim_{x\rightarrow y} \langle \phi(x)\phi(y) \rangle \bar{l^2} \nonumber \\
  &=& \lim_{x\rightarrow y} \frac{L_p^2}{4\pi^2}\cdot
  \frac{1}{(x-y)^2} \bar{l^2} = \frac{L_p^2}{4\pi^2}
  \label{EHPlanckLength}
\end{eqnarray}
In other words, quantum fluctuations produce a ``ground state
length'' just as a harmonic oscillator has a ground state energy.

Note that the path integral approach taken here obscures the fact
that the conformal factor is not a true dynamical field subject to
quantum fluctuations. The source of this confusion is the apparent
kinetic term in the action \eref{EHConformalAction}, which
justifies all subsequent steps leading to
\eref{EHPlanckLength}. However, in the hamiltonian framework, the
trace part of the metric perturbations does not have a canonically
conjugate momentum, and a true kinetic term for the conformal
factor should not appear in the action.

The explanation for the offending term is hidden in the measure of
\eref{PathIntegral} and was finally resolved by Mazur and Mottola
\cite{Mazur1990Tp}.  To identify the correct measure, they first
decomposed the space of metric perturbations into diffeomorphisms and
physical fluctuations.  The remaining physical subspace was further
decomposed into constrained (conformal) and dynamical
(transverse-traceless) degrees of freedom. Seen in this light,
\eref{FullMetric} amounts to a change of coordinates in the space
of metrics, which introduces a non-trivial Jacobian in the measure.
A field redefinition of the conformal factor then turns the apparent
kinetic term in \eref{EHConformalAction} into a potential term,
confirming the result that the conformal modes are non-propagating
constrained modes.

\section{A Dynamical Conformal Field}
While the conformal factor is non-propagating in pure
Einstein-Hilbert gravity, the classical constraints that fix the
conformal part of the metric fluctuations in terms of matter sources
cannot be maintained upon quantization \cite{Antoniadis1992Cs}.  The
trace anomaly of matter coupled to gravity induces an effective
action for the conformal factor that gives rise to non-trivial
dynamics \cite{Antoniadis19924d}.  In other words, the conformal
factor is promoted to a dynamical field when gravity is coupled
to quantized matter. Thus we can revisit Padmanabhan's calculation
in light of this dynamical model of the conformal factor.

We begin by summarizing the basic results of Antoniadis, Mazur and
Mottola \cite{Antoniadis1992Cs}.  The effective action of the
conformal factor becomes local in the conformal parameterization
\begin{equation} g_{\mu \nu}(x) =
e^{2\sigma(x)}\bar{g}_{\mu \nu}(x),
\end{equation}
where $\bar{g}_{\mu \nu}$ is a fiducial metric.  The total
effective action is
\begin{equation}
  S = S_{\mathrm{EH}} + S_{\mathrm{matt}} + S_{\mathrm{anom}},
  \label{EffectiveAction}
\end{equation}
where $S_{\mathrm{EH}}$ is the Einstein-Hilbert action \eref{EHAction} evaluated at $g=e^{2\sigma}\bar{g}$,  $S_{\mathrm{matt}}$ is the action for matter fields, and
$S_{\mathrm{anom}}$ is the trace anomaly-induced effective action
\cite{Riegert1984}
\begin{equation}
  S_{anom}[\bar{g};\sigma] = \int \rmd^4x \sqrt{-\bar{g}}
      \left[2b^\prime \sigma \bar{\Delta}_4 \sigma + b^\prime \left(\bar{E} - \frac{2}{3} \bar{\Box}
      \bar{R}\right)\sigma + b\bar{F}\sigma  \right]. \label{SAnom}
\end{equation}
Here, $\Delta_4$ is the conformally invariant fourth-order operator
\begin{equation}
  \Delta_4 = \Box^2 + 2R^{\mu \nu}\nabla_\mu \nabla_\nu - \frac{2}{3} R \Box
  + \frac{1}{3}(\nabla^\mu R)\nabla_\mu
\end{equation}
and
\begin{eqnarray}
  F\equiv C_{\mu \nu \rho \lambda}C^{\mu \nu \rho \lambda}=R_{\mu \nu \rho \lambda}R^{\mu \nu \rho \lambda}-2R_{\mu \nu}R^{\mu \nu} +
  \frac{1}{3}R^2\\
  E\equiv R_{\mu \nu \rho \lambda}R^{\mu \nu \rho \lambda} - 4R_{\mu \nu}R^{\mu \nu} + R^2
\end{eqnarray}
are the square of the Weyl tensor and the Gauss-Bonnet integrand,
respectively.  The coupling constants $b$ and $b^\prime$ depend on
the matter content of the theory \cite{Antoniadis1992Cs,Duff1977}:
\begin{eqnarray}
  b &=& \frac{1}{16\pi^2}\frac{1}{120}(N_S + 3N_F + 12N_V - 8) +
  b_{\mathrm{grav}}\\
  b^\prime &=& -\frac{1}{32\pi^2}Q^2 \nonumber\\
  &=& -\frac{1}{16\pi^2}\frac{1}{360}\left(N_S + \frac{11}{2}N_F +
  62N_V - 28\right) + b^\prime_{\mathrm{grav}},
  \label{CentralCharge}
\end{eqnarray}
where $N_S$, $N_F$ and $N_V$ are the numbers of scalar, Weyl
fermion, and vector fields.  The spin-0 and ghost
contributions are included in the -8 and -28 factors, while
$b_{\mathrm{grav}}$ and $b^\prime_{\mathrm{grav}}$ count the
contributions from the spin-2 metric fields. Because the values of these
gravitational contributions, as well as contributions beyond the
Standard Model, remain open questions, $Q^2$ will be treated as a
free parameter.

The total trace anomaly of the full theory described by \eref{EffectiveAction} must vanish \cite{Antoniadis1992Cs}.  The absence of this anomaly requires that the vacuum is a conformal fixed point at which the $\beta$ functions of all couplings must vanish.  The physical metric then acquires an anomalous scaling dimension
\begin{equation}
  g_{\mu \nu}(x) = e^{2\alpha \sigma(x)}\bar{g}_{\mu \nu}(x),
\end{equation}
where $\alpha$ is determined by the $\beta$ function for the Einstein-Hilbert action \cite{Antoniadis19924d},
\begin{equation}
  \alpha = \frac{1-\sqrt{1-\frac{4}{Q^2}}}{\frac{2}{Q^2}}. \label{alpha}
\end{equation}

From here we can follow Padmanabhan's prescription.  Looking only at
conformal fluctuations and choosing a Minkowski fiducial metric
$\bar{g}_{\mu \nu} = \eta_{\mu \nu}$, the action
\eref{EffectiveAction} reduces to
\begin{equation}
  S_{eff}[\sigma] = -\frac{Q^2}{(4\pi)^2}\int \rmd^4x \sigma
  \bar{\Box}^2 \sigma + \frac{1}{8\pi G}\int \rmd^4x \left[3e^{2\alpha \sigma}\left(\partial_a \sigma\right)^2 - \Lambda
  e^{4 \alpha \sigma}\right]. \label{ConformalEffective}
\end{equation}
The action simplifies again by
invoking the translational invariance of the measure and shifting
$\sigma$ by a constant $\sigma_0$ \cite{Antoniadis19924d}. In the
limit $\sigma_0 \rightarrow -\infty$, the final terms drop out,
leaving only the free quartic action. The propagator for this
fourth-order kinetic term is $k^{-4}$ in momentum space, which is
just a logarithm in coordinate space:
\begin{equation}
  \langle \sigma(x) \sigma(y) \rangle =
  -\frac{1}{2Q^2} \ln [\mu^2 (x-y)^2], \label{propagator}
\end{equation}
where $\mu$ is an infrared cutoff.

Now the expectation value of the interval \eref{Interval} becomes
\begin{eqnarray}
  \lim_{x\rightarrow y} \langle \rmd s^2(x,y) \rangle &=& \lim_{x\rightarrow y}
      \langle e^{\alpha \sigma(x)}e^{\alpha \sigma(y)} \rangle \eta_{\mu \nu}\rmd x^\mu \rmd x^\nu \nonumber\\
  &=& \lim_{x\rightarrow y} e^{\alpha^2 \langle \sigma(x)\sigma(y) \rangle}
      \bar{\ell^2}(x,y) \nonumber \\
  &\propto & \lim_{x\rightarrow y} \left[\bar{\ell}(x,y)\right]^{2-\alpha^2/2Q^2}, \label{metricexpectation}
\end{eqnarray}
where the first line makes use of covariant point-splitting, and normal ordering is used on each operator $e^{\alpha \sigma (x)}$ individually.  The second equality in \eref{metricexpectation} is a standard field theory result that makes use of the Baker-Campbell-Hausdorff relation\footnote{This result requires that the operators in the exponent be no more than linear in creation/annihilation operators, and that the creation and annihilation operators obey standard commutation relations.  While this is certainly true for a free Klein-Gordon field, it is no longer obvious for the quartic action \eref{ConformalEffective}.  For example, we expect a quartic field to have two sets of creation and annihilation operators.  Recent efforts to quantize the conformal factor in $R \times S^3$ \cite{Antoniadis1997} and Minkowski space \cite{Hamada2011} confirm both of these requirements.}.  This yields the interesting result that the scaling depends on the matter content:  the distance approaches zero for all values $Q^2 < -1/12$ or $Q^2 \geq 4$.  From \eref{alpha}, positive values of $Q^2<4$ are excluded at the conformal fixed point. The interval is constant at the critical point $Q^2 = -1/12$, and $-1/12 < Q^2 < 0$ gives the nonsensical result that distances diverge in the limit $\bar{\ell}\rightarrow 0$.  For large $Q^2$, the interval scales as $2-\frac{1}{2Q^2}$, and classical scaling is recovered in the limit $|Q^2|\rightarrow \infty$.

It follows from \eref{CentralCharge} that $Q^2>0$ for normal
matter; however, it is worth noting that some models of conformal
supergravity contribute negatively to $Q^2$ \cite{Fradkin1985}.
Calculations of the one-loop contributions from Einstein
gravity place it at $Q^2_{grav} \approx 7.9$
\cite{Christensen1980,Antoniadis1992Cs}. Together, the Standard
Model particle content ($N_F=45$ and $N_V = 12$) and one-loop
gravitational contributions give a value
\begin{equation}
  Q^2_{\mathrm{SM}} \approx 13.2.
\end{equation}
The greatest uncertainty in the value of $Q^2$ comes from the gravitational contributions, and a precise theoretical prediction for $Q^2$ remains an open problem.  Recent attempts to place observational limits on $Q^2$ using WMAP
data claim to limit $Q^2$ to the range $|Q^2|>80$ \cite{Antoniadis2007}.

Thus a more complete treatment of conformal fluctuations using the
trace anomaly-induced effective action do not place a lower bound on
the distance between two points.  Of course this result should be
viewed with some skepticism.  In particular, the spin-2 metric fluctuations are expected to become important around the Planck scale but have been frozen out in this approach.  Additionally, the transition from Einstein gravity to the conformally invariant phase described by \eref{SAnom} is poorly understood, and more research is need to determine the scales at which the effective action becomes significant.

\ack
I would like to thank Steve Carlip for comments on early versions of this paper.  This work was supported in part by the Department of Energy grant DE-FG02-91ER40674.

\end{document}